\documentclass[modadp,a4paper,fleqn]{modw-art}
\usepackage{times,cite,modw-thm}
\usepackage{times,cite}
\usepackage[]{graphicx}

\newcommand{\beq}{\begin{equation}}
\newcommand{\eeq}{\end{equation}}
\newcommand{\beqa}{\begin{eqnarray}}
\newcommand{\eeqa}{\end{eqnarray}}

\def\la{\lower.5ex\hbox{$\; \buildrel < \over \sim \;$}}
\def\ga{\lower.5ex\hbox{$\; \buildrel > \over \sim \;$}}

\begin{document}

\DOIsuffix{theDOIsuffix}
\Volume{16}
\Month{03}
\Year{2012}
\pagespan{1}{}



\title[Evanescent Matter]{Evanescent Matter}

\author[P. J. E. Peebles]{P. J. E. Peebles
  \footnote{Corresponding author E-mail:~\textsf{pjep@Princeton.edu},
            Phone: 609\,258\,4386}}
\address[]{Joseph Henry Laboratories, Princeton University, Princeton, NJ 08544, USA}
\begin{abstract}
The $\Lambda$CDM cosmology offers a picture for galaxy formation that is broadly promising but difficult to reconcile with the evidence that galaxies were assembled earlier than seems naturally to follow from this  cosmology, and that environment has had strikingly little effect on the evolution of ellipticals and pure disk spiral galaxies after assembly. Reconciliation might be aided by adding to  $\Lambda$CDM evanescent matter that has an evolving mass and a fifth force large enough to aid earlier assembly of more nearly isolated protogalaxies.
\end{abstract}

\maketitle

\section{Introduction}

The hierarchical growth of cosmic structure by merging and accretion is predicted by the standard $\Lambda$CDM cosmology, merging is seen in disturbed galaxies and inferred for close pairs of galaxies that intuition and theory argue will  suffer mergers \cite{mergers}, and it is a basis for analyses of galaxy formation that can account for a broad range of observations (e.g. \cite{Ostriker,Governato} and references therein). Despite these successes the phenomenology in Section~\ref{sec:phenomenology} suggests that a still better cosmology would predict earlier assembly of the large galaxies, with suppression of the merging and accretion that is such a prominent feature of $\Lambda$CDM but has had such a subtle effect on elliptical and pure disk spiral galaxies. The addition to $\Lambda$CDM proposed in Section~\ref{sec:ev} offers a way to resolve the apparent discrepancy between theory and observation. It couples a scalar or pseudoscalar field, perhaps one suggested by superstring pictures (e.g. \cite{PeebGub,Dimopoulos,Marsh-etal,DienesThomas} and references therein), to a spin-1/2 evanescent matter particle by the familiar Yukawa interaction. The simple illustrations in Section~\ref{AnIllustration}, in spherical accretion and linear perturbation theory,  show how evanescent matter can promote earlier formation of nonlinear mass concentrations, at least roughly in line with the evidence in Section~\ref{sec:phenomenology}, while having little effect on structure on scales large compared to galaxies.   

 \section{Phenomenology}\label{sec:phenomenology}

Here are four interesting challenges to the current paradigm for structure formation.\medskip

\noindent (1) {\it Pure Disk Galaxies.} In a pure disk galaxy such as the Milky Way most of the visible stars are largely rotationally supportted in a disk or bar \cite{KK,Shenetal}. This does not seem to be an uncommon situation: Kormendy {\it et al.} \cite{Kormendyetal} find that 11 of the nearest 19 large galaxies (with circular velocities $v_c>150$ km~s$^{-1}$) are pure disks. In these galaxies stars had to have formed largely after diffuse baryons had settled onto the growing disk. If the galaxy were assembled by merging of subhalos that already contained stars then the early generations of stars would form a stellar halo or classical stellar bulge. Successful models for galaxies with small bulges place a tight limit on the critical density for star formation, which has the effect of suppressing star formation in the merging subhalos and delaying star formation in the growing disk to redshift $z \la 2$ (\cite{Madau,Moore}). But it is curious that star formation in this apparently common class of large galaxies must be supposed to have commenced well after the peak of the global star formation rate. 

The pure dark matter Aquarius \cite{Springel} halos selected for conditions resembling that of the Milky Way indicate that the matter now in the luminous part of a large galaxy was at redshift $z\sim 3$ in subhalos spread over $\sim 100$\,kpc (as illustrated in Fig. 5 in \cite{PeebVenice}). Many of these subhalos were dense enough for the baryons to  cool and collapse in a Hubble time, and the gravitational potentials were deep enough to bind baryons heated by stellar photoionization (as analyzed in unpublished work with Jie Wang and Adi Nusser). It is difficult to see why stars would not form in these subhalos. If they did it could account for the classical bulge in M\,31, and for the stars in dwarf galaxies, but it could not account for pure disk galaxies such as the Milky Way. 

Shen, Rich, Kormendy {\it et al.}  \cite{Shenetal} conclude that pure disk galaxies ``present an acute challenge to the current picture of galaxy formation in a universe dominated by cold dark matter---growing a giant galaxy via hierarchical clustering ($V_c\simeq 220$ km s$^{-1}$ in the Milky Way) involves so many mergers that it seems almost impossible to avoid forming a substantial classical bulge.'' Rationalizing this fascinating conflict between theory and observation might be aided by adjusting $\Lambda$CDM to promote assembly of pure disk galaxies  before the peak of global star formation, leaving fewer remnants to be accreted when stars were forming in abundance. 
\medskip

\noindent (2) {\it Scaling Relations for Early-type Galaxies.} The preference of the most massive early-type galaxies for the densest environments \cite{Dressler1980} follows in a natural way from the hierarchical assembly by mergers predicted by the $\Lambda$CDM cosmology. Since the rate and nature of mergers depends on the ambient density it also is very natural to expect that the properties of ellipticals depend on the environment. This does not agree with the remarkable insensitivity of the general properties of ellipticals to their environment. 

The spectra of elliptical galaxies correlate with the stellar velocity dispersion $\sigma$: ellipticals with larger $\sigma$ are redder. This correlation is strikingly insensitive to the abundance of neighboring galaxies (as discussed in \cite{Hogg,Zhu}, and illustrated in Fig. 5 in \cite{islanduniverse}). The physical conditions and processes of evolution that account for the observed variations of spectra are under discussion \cite{GravesFaber}. The key point for our purpose is that the different conditions of different elliptical galaxies produced differences of spectra that correlate with a measure of internal structure, $\sigma$, as one might expect, but are much less sensitive to a measure of external conditions, the ambient density, which was not to be expected under $\Lambda$CDM. This could be read to mean (a) variations in environment affect both spectrum and $\sigma$, but the effect is to shift ellipticals along the spectrum-$\sigma$ relation, or (b) environment does not much matter because ellipticals were assembled at high redshift, when there was not much variation in environment, followed by near passive evolution (or more generally evolution in isolation from the surroundings). The same line of argument applies to the insensitivity to environment of the relation between red galaxy mass-to-light ratio and radius \cite{Bernardi,Magoulas}, the relation between elliptical galaxy luminosity and radius \cite{Nair}, and the relation between galaxy stellar mass and radius \cite{Maltby}. These several observations argue against the accidental interpretation (a). Interpretation (b) does not seem consistent with the late-time merging in the illustration in Figure~5  \cite{PeebVenice}) from the Aquarius simulations. Again, rationalizing the situation might be aided by  galaxy assembly earlier than predicted by $\Lambda$CDM, when differences in environment were less pronounced. \medskip

\noindent (3) {\it Expanding Spheroids.}  Massive apparently quiescent early-type galaxies at redshifts $z\sim 2$ are expanding at rates larger than would be expected from the effects of star formation \cite{vanDokkum2010}, and perhaps even from the effects of dry mergers \cite{Ellis}. This behavior is made more interesting by point (2), the striking insensitivity of the present properties of ellipticals to the present environment. Although the effect of environment on the relation between elliptical galaxy luminosity and radius depends on details of how radii are defined, the effect is small in any case \cite{Nair}, and not what one would have expected. Nair, van den Bergh and Abraham \cite{Nair} conclude that these observations ``challenge the plausibility of the merger-driven hierarchical models for the formation of massive ellipticals.'' Similar arguments may be applied to the other scaling relations for early-type galaxies mentioned in point (2). This interesting situation might point to the non-dissipative loss of evanescent matter mass in the  picture discussed in the next section. 

\medskip\noindent (4) {\it Satellites of the Milky Way.} The large number of predicted dark matter halos in the Local Group compared to the number of observed galaxies  \cite{Klypin, Moore99} has been widely discussed. An important recent advance \cite{Bullock} makes use of the internal velocities and radii of the brighter satellites of the Milky Way, which are compared to what would be expected from the $\Lambda$CDM Aquarius halos that are meant to model the situation of the Milky Way \cite{Springel}. The results lead  Boylan-Kolchin, Bullock, and Kaplinghat \cite{Bullock} to conclude that ``the most massive subhalos in galaxy-mass dark matter (DM) haloes in $\Lambda$ cold dark matter ($\Lambda$CDM) are grossly inconsistent with the dynamics of the brightest Milky Way dwarf spheroidal galaxies.'' Yet again, the way out of this challenging situation might involve more rapid structure formation that more thoroughly removed or puffed up the more compact subhalos by merging with each other or the host.  

I cannot forebear also mentioning  the curious properties of the Local Void. This low density region is bounded on one side by the plane of the Local  Supercluster, which contains the Local Group and several other nearby groups. Two large spiral galaxies, M\,101 and NGC\,6946, each accompanied by the usual satellites, are in islands close to the near edge of the Local Void. Apart from that, the Local Void occupies about one third of the volume within 8 Mpc distance (as illustrated in Fig. 1 in \cite{PeebNusser}), but among the $\sim 562$ identified galaxies within 8 Mpc only three dwarf galaxies are in the one third of the volume of this part of the Local Void and outside the islands of satellites around M\,101 and NGC\,6946. A recent analysis \cite{void-simulations} of void formation in $\Lambda$CDM indicates that the segregation of galaxies by mass is a smoothly varying function of distance into a void. This is not suggested by what is known about the abrupt edge of the Local Void along the Local Supercluster. Kreckel, Joung and Cen \cite{void-simulations} conclude that ``we are not compelled to suggest that any alteration of the current standard CDM paradigm is required'' \cite{void-simulations}. This agrees with the earlier analysis of Tinker and Conroy \cite{TinkerConroy}, and the conclusion certainly is fair. I continue to suspect that the properties of the Local Void are curious enough to merit attention, however, perhaps even indicating that the structure formation that assembled galaxies somewhat earlier than suggested by $\Lambda$CDM may also have more completely emptied voids. 

The theory of how galaxies formed certainly must take account of very significant departures from an island universe picture. For example, S0 galaxies offer a persuasive case for the influence of environment on spiral galaxies \cite{S0s,KormendyBender}.  But the existence of the several challenges mentioned here, based on a considerable variety of lines of evidence, but with a common indication of discrepancies with the description of merging and evolution in $\Lambda$CDM, is interesting enough to motivate considerations of possible adjustments to the present paradigm. 

\section{Evanescent Matter Model}\label{sec:ev}

This proposed adjustment adds to $\Lambda$CDM the Lagrangian density 
\beq
L_{\rm ev} = i\bar\psi\gamma_\mu\partial_\mu\psi - \bar\psi (\lambda_1\phi_1 + i\gamma_5\lambda_2\phi_2)\psi + 
{1\over 2}\partial^\mu\phi_1\partial_\mu\phi_1 + {1\over 2}\partial^\mu\phi_2\partial_\mu\phi_2. \label{eq:action}
\eeq
The field $\psi$ describes a spin-1/2 Dirac particle with Yukawa coupling to a scalar field $\phi_1$ and pseudoscalar $\phi_2$ with constants $\lambda_1$ and $\lambda_2$. Though these components interact only with gravity it will avoid confusion to term this evanescent matter, reserving the name dark matter for the standard component of $\Lambda$CDM. 

Scalar interactions similar to what is described in Eq.~(\ref{eq:action}) have been under discussion for a long time (in literature reviewed in \cite{FarrarPeebles}; later examples are in \cite{ClampittJainKhoury,Tarrant}). The direction taken in some recent discussions (e.g. \cite{PeebGub,Dimopoulos,Marsh-etal,DienesThomas} and references therein) is motivated by the tendency of superstring pictures to provide scalar and pseudoscalar fields with masses that may be comparable to or even smaller than Hubble's constant (in the units used here, $\hbar=1=c$).  Whether this or something else might be the provenance of the scalar and pseudoscalar fields in $L_{\rm ev}$, the fields offer the familiar Yukawa interaction to a spin-1/2 particle, which may be chiral, as in the standard model for particle physics, but here without gauge symmetry or a Higgs arrangement to fix the  field values. The possibly new idea presented here is that this evanescent mass component has nothing to do with the dark matter or dark energy of $\Lambda$CDM, or with completion of the gravity theory. It is instead a component of matter added to $\Lambda$CDM that could have  interesting effects on young galaxies.

If, as will be assumed, $\phi_1$ and $\phi_2$ evolve on very long time scales, the $\psi$ particles behave as if they had a mass
\beq
m = \sqrt{m_1^2 +m_2^2}, \quad m_1=\lambda_1\phi_1, \quad m_2= \lambda_2\phi_2, \label{eq:the-mass}
\eeq
where as usual one takes the positive square root. This follows from the chiral rotation
\beq
\psi = e^{-i\gamma_5\theta/2}\psi^\prime, \qquad \tan\theta = m_2/m_1,
\eeq
which brings the action expressed in terms of $\psi^\prime$ to the usual Dirac form with the particle mass $m$ in Eq.~(\ref{eq:the-mass}).

It is also assumed that the particle de Broglie wavelengths are much shorter than any other scale of interest, and the scalar and pseudoscalar field values are large, so the matter may be described as a gas of  classical particles that interact with classical fields $\phi_1$ and $\phi_2$ in the action  
\beq
S_{\rm ev} =\int d^4x\sqrt{-g}\,(\partial^\mu\phi_1\partial_\mu\phi_1 +
\partial^\mu\phi_2\partial_\mu\phi_2 )/2 - 
\sum _i\int  m_i\,ds_i,  \label{eq:classical-2-particle-acton}
\eeq
where $m_i$ is the mass in Eq.~(\ref{eq:the-mass}) at the position of particle $i$. 

It is worth pausing to note another way to see the origin of the evanescent matter particle mass $m_i$. An exercise in gamma matrices shows that the solution to the single particle Dirac equation in flat spacetime, 
\beq
i\gamma^\mu\partial_\mu\psi = (m_1+i\gamma_5m_2)\psi,
\eeq
for a plane wave normalized to $\psi^\dagger\psi=1$, yields
\beq
\bar\psi\psi = \sqrt{1-v^2}m_1/m, \quad i\bar\psi\gamma_5\psi = \sqrt{1-v^2}m_2/m,
\eeq
where the particle speed is $v$. The potential energy term in Eq. (\ref{eq:action}) for a single particle thus  is
\beq
\bar\psi (\lambda_1\phi_1 + i\gamma_5\lambda_2\phi_2)\psi \rightarrow 
\sqrt{1-v^2}(m_1^2 + m_2^2)/m = m\sqrt{1-v^2}. 
\eeq
The reciprocal Lorentz factor in this expression corresponds to the scalar line interval in the classical particle action in the last term on the right hand side of Eq.~(\ref{eq:classical-2-particle-acton}). 

It might also be noted that since the model is applied in the classical particle limit the Dirac field $\psi$ in Eq.~(\ref{eq:action}) can be replaced by a scalar field $\chi$ with potential $(\lambda_1^2\phi_1^2+\lambda_2^2\phi_2^2)\chi^2/2$. This gives evanescent particle mass $m =\sqrt{ \lambda_1^2\phi_1^2+\lambda_2^2\phi_2^2}$, as before, and as before the theory admonishes us to take the positive square root.

The mass $m$ will be taken to be large at high redshift, comparable to the Planck mass. The field values are attracted toward zero because that minimizes the energy $m$, meaning the particle masses  are decreasing, or evanescent, though not through any dissipative process. The value of $m$ decreases more rapidly in places where  there are more evanescent particles, and the resulting spatial gradient of  $m$ produces a fifth force of attraction among evanescent particles. The particle momentum changes in response to gravity and the fifth force, meaning that decreasing $m$ increases the particle velocity.  The result of this effect and the fifth force is that departures from homogeneity grow much more rapidly in the evanescent matter than in the dark matter. This means the evanescent matter can develop strongly nonlinear mass concentrations that dominate the local mass density even if the mean mass density in evanescent matter is subdominant. These evanescent matter concentrations, which tend to be transient, can gravitationally attract local concentrations of dark matter that form earlier than in standard $\Lambda$CDM. This is in the direction suggested by the phenomenology reviewed in Section~\ref{sec:phenomenology}.

\section{Illustrations: spherical accretion and linear perturbation theory}\label{AnIllustration}

This simplified version assumes a single field, with $\lambda_2=0$ in Eq.~(\ref{eq:classical-2-particle-acton}). (This is equivalent to $\lambda_1=\lambda_2$ if the strongly decaying mode is suppressed in both fields, for then $\phi_1/\phi_2$ is constant.) Here the classical action  is 
\beq
S_{\rm ev} =\int d^4x\sqrt{-g}\,(\partial\phi_1)^2/2 - 
\sum _i\int  |\phi_1(\vec x_i,t)|\,ds_i, \label{eq:singleclassicalaction}
\eeq
where the path of evanescent particle $i$ is $x_i(t)$. As discussed in connection with Eq.~(\ref{eq:the-mass}), the absolute value of the field enters the path integral. The constant $\lambda_1$ has been scaled to unity in this form for $S_{\rm ev}$ by scaling the particle number density (in the sum over particles $i$). We can take it that spacetime curvature fluctuations are small and described by the Newtonian gravitational potential ${\cal U}$, so
\beq
ds_i/dt = \sqrt{1 + 2\,{\cal U} - (1 - 2\,{\cal U})a^2(dx_i/dt)^2}. \label{eq:singleclassicalaction1}
\eeq
It will be necessary to consider relativistic particle motions when $\phi$ passes trough zero, but for simplicity that discussion is deferred to Section~\ref{sec:accretion} (Eq.~\ref{throughzero}). To the nonrelativistic order $v^2$ we can rewrite Eqs.~(\ref{eq:singleclassicalaction}) and~(\ref{eq:singleclassicalaction1}) as the Lagrangian density
\beq 
{\cal L}_{\rm ev} ={\dot\phi_1^2\over 2} - {\nabla\phi_1^2\over 2a^2} - 
   \sum_i{\delta^3(\vec x - \vec x_i(t))\over a^3} |\phi_1(\vec x,t)|\left(1 + \,{\cal U}(\vec x,t) - 
   {1\over 2}a^2\dot x_i^2\right). \label{eq:lagrangian}
\eeq
The field equation from this Lagrangian is
\beq
\ddot\phi_1(t)+3{\dot a\over a}\dot\phi_1 - {\nabla^2\phi_1\over a^2} = - n(\vec x,t)\,{\rm sgn}(\phi_1),
\label{eq:Phieq}
\eeq
where $n(\vec x,t)$ is the proper number density of  evanescent particles, and ${\rm sgn}(\phi)=1$ if $\phi\geq 0$, ${\rm sgn}(\phi)=-1$ if $\phi< 0$. With 
\beq
\phi_1(\vec x, t) = \phi(t)  + \varphi(\vec x,t)\, {\rm sgn}(\phi),
\eeq
the homogeneous part of the field equation is
\beq\ddot\phi(t)+3{\dot a\over a}\dot\phi = -\bar n(t)\, {\rm sgn}(\phi),  \label{eq:meanphi}
\eeq
and the inhomogeneous part is
\beq
\ddot\varphi(\vec x,t)+3{\dot a\over a}\dot\varphi - {\nabla^2\varphi\over a^2} =
    - (n(\vec x,t) - \bar n(t)).  \label{eq:phifield}
\eeq

The nonrelativistic evanescent matter particle equation of motion from Eq.~(\ref{eq:lagrangian}) is
\beq
{d \over dt}a^2|\phi|{d\vec x\over dt} = -|\phi|\nabla\,{\cal U} - \nabla\varphi. 
\label{eq:eqom}
\eeq
Since $\varphi$ typically is small compared to $\phi$  the mean value $|\phi(t)|$ serves as the particle mass in the momentum, $a^2|\phi| d\vec x_i/dt$, and in the gravitational force, $-|\phi|\nabla\,{\cal U}$.  The gradient of the mass in the last term in this equation of motion is the fifth force. The treatment of zero crossings of $\phi$, where the motion of an evanescent particle is transiently relativistic, is discussed in Section~\ref{sec:accretion}.

If the length scale $\ell$ of a density fluctuation $n-\bar n$ is much smaller than the expansion time $t$ then the time derivatives in Eq.~(\ref{eq:phifield}) are subdominant to the space derivatives, and the equation simplifies to
\beq
 \nabla^2\varphi/ a^2 = n(\vec x,t) - \bar n(t), \label{eq:newphifield}
\eeq
which can be compared to the Newtonian gravity equation
\beq
\nabla^2\,{\cal U}/a^2 = 4\pi G(\rho-\bar\rho) = 4\pi G[\rho_{\rm dm} - \bar\rho_{\rm dm} + |\phi|\,(n - \bar n)]. \label{eq:gravpoisson}
\eeq
Thus at $\ell\ll t$ the ratio of the fifth force $-\nabla\varphi$ to the gravitational force $-|\phi|\nabla\,{\cal U}$ of interaction between evanescent matter particles is 
\beq
f_{\rm ev}/f_{\rm g} = 1/(4\pi G\phi^2).  \label{eq:forceratio}
\eeq
This is the square of the ratio of the Planck mass (suitably normalized) to the evanescent particle mass. 

In the other limit, $\ell \gg t$, the space derivatives in Eq.~(\ref{eq:phifield}) are subdominant to the time derivatives and the fifth force is suppressed by order $(t/\ell)^2$. It will be recalled that the inverse square law for gravity also applies when $\ell > t$, where the fifth force is suppressed.

\subsection{Parameters}\label{sec:parameters}

The parameters
\beq
F_{\rm ev} = {1\over 4\pi G\phi_i^2}, \qquad R_{\rm ev} ={ \bar n\phi_i\over\bar\rho_{\rm dm}}. 
\label{eq:parameters}
\eeq 
represent the primeval values (at $a\rightarrow 0$) of the field $\phi_i>0$,  the strength $F_{\rm ev}$ of the fifth force relative to gravity (Eq.~(\ref{eq:forceratio})) for the interaction of evanescent particles, and the ratio $R_{\rm ev}$  of evanescent matter mass density to dark matter mass density $\bar\rho_{\rm dm}$. We can simplify the equation of motion of the dark matter particles by scaling the dark matter mass $m$ to the same mean number densities of dark and evanescent particles, $\bar n =\bar n_{\rm dm}$. Then in the notation of Eq.~(\ref{eq:parameters}) the nonrelativistic equations of motion (\ref{eq:eqom}) are
\beqa
&&{1\over a}{d \over dt}a^2{d\vec x_i\over dt} = {Gm\over a^2}
 \left[\sum_{j,\,{\rm dm}}{\vec x_j-\vec x_i\over x_{ji}^3}+R_{\rm ev}{|\phi|\over\phi_i }
   \sum_{l,\,{\rm ev}}{\vec x_l-\vec x_i\over x_{li}^3} \right],  \nonumber \\
&& \hspace{-10mm}{1\over a|\phi|}{d \over dt}a^2|\phi|{d\vec x_k\over dt} = {Gm\over a^2}
 \left[\sum_{j,\,{\rm dm}}{\vec x_j-\vec x_k\over x_{jk}^3}+R_{\rm ev}\left({|\phi|\over\phi_i } +
 F_{\rm ev}{\phi_i\over|\phi |} \right)\sum_{l,\,{\rm ev}}{\vec x_l-\vec x_k\over x_{lk}^3} \right],
\label{eq:particle-eqom}
\eeqa
for dark and evanescent matter. It will be recalled that the sums must be ordered to converge to zero when the particle distributions are homogeneous. The inverse square law in the fifth force in the second line assumes convergence is reached on scales small compared to the Hubble length $t$. 

A first integral of  Eq.~(\ref{eq:meanphi}) is
\beq
\dot\phi = -\bar n \int_0^t dt\,{\rm sgn}(\phi). \label{eq:phidot}
\eeq
I have set the constant of integration to zero to avoid the singularity behavior of $\phi$ at $a(t)\rightarrow 0$. With the time unit $t_e$ defined by 
\beq
{1\over t_e^2} = {4\over 3}\pi G\bar\rho_{\rm dm}(z_{\rm eq})={4\over 3}\pi G\bar\rho_{\rm dm}(t)a(t)^3,
\label{eq:timeunit}
\eeq
where $\bar\rho_{\rm dm}(z_{\rm eq})$ is the dark matter density at the redshift $z_{\rm eq}$ at equal mass densities in radiation and dark matter (here including baryons), and $a=1$ at $z_{\rm eq}$, the result of integrating Eq.~(\ref{eq:phidot}) may be expressed as 
\beq
\phi/\phi_i = 1 - 3R_{\rm ev}F_{\rm ev}\int _0^t dt'\, a(t')^{-3}\int _0^{t'} dt^{\prime\prime}{\rm sgn}(\phi(t^{\prime\prime})).  \label{eq:pi-iint}
\eeq
Before $\phi$ has first passed through zero, and at $a\ll a_{\rm eq}$, this is 
\beq
\phi(t)/\phi_i = 1 - {3\over 4}R_{\rm ev}F_{\rm ev}a(t)/a_{\rm eq}. \label{eq:phiint}
\eeq
If $R_{\rm ev}F_{\rm ev}\ll 1$ then $\phi$ has not changed much prior to $z_{\rm eq}$, and the subsequent evolution before $\phi$ passes through zero and before $\Lambda$ becomes important is 
\beq
\phi(t) /\phi_i \sim 1 - R_{\rm ev}F_{\rm ev}\log (a(t)/a_{\rm eq}). \label{eq:phioft}
\eeq
One sees that $R_{\rm ev}F_{\rm ev}$ must be tuned to a value slightly less than unity to make $\phi$ first pass through zero and the evanescent matter do something interesting during the early  assembly of protogalaxies. The examples in the next section show there is more freedom in the choice of the ratio $R_{\rm ev}$ of primeval mass densities in evanescent and dark matter, with larger $R_{\rm ev}$ requiring  a smaller measure $F_{\rm ev}$ of the fifth force, but producing a larger effect on the growth of concentrations of  dark matter by the transient concentrations of the evanescent matter mass. 

In the time unit of Eq.~(\ref{eq:timeunit}) the Friedmann-Lem\^\i tre equation is
\beq
{1\over 2}\left( da\over dt\right)^2 = {1\over a^2} + {1+R_{\rm ev}|\phi|/\phi_i\over a} + 
 {\Omega_\Lambda a^2\over a_o^3\Omega_{\rm dm}}  + {3\over 2}{R_{\rm ev}^2F_{\rm ev}\over a^4}\left(\int_0^t dt\,{\rm sgn}(\phi) \right)^2.
\label{eq:FLeq}
\eeq
The first source term is the contribution by the cosmic thermal background radiation. The second term represents the sum of mass densities in conventional and evanescent matter. In the third term  $\Omega_\Lambda$ and  $\Omega_{\rm dm}$ are the fractional contributions to Hubble's constant by the cosmological constant and dark matter (including baryons) and $a_o = 3230$ is the redshift at equality of mass densities in radiation and conventional matter. The last term represents the field energy density $\dot\phi/2$ (Eq.~(\ref{eq:phidot})). It is small in the numerical examples, but included in the computations.

 \begin{figure}[htpb]
\begin{center}
\includegraphics[angle=0,width=2.5in]{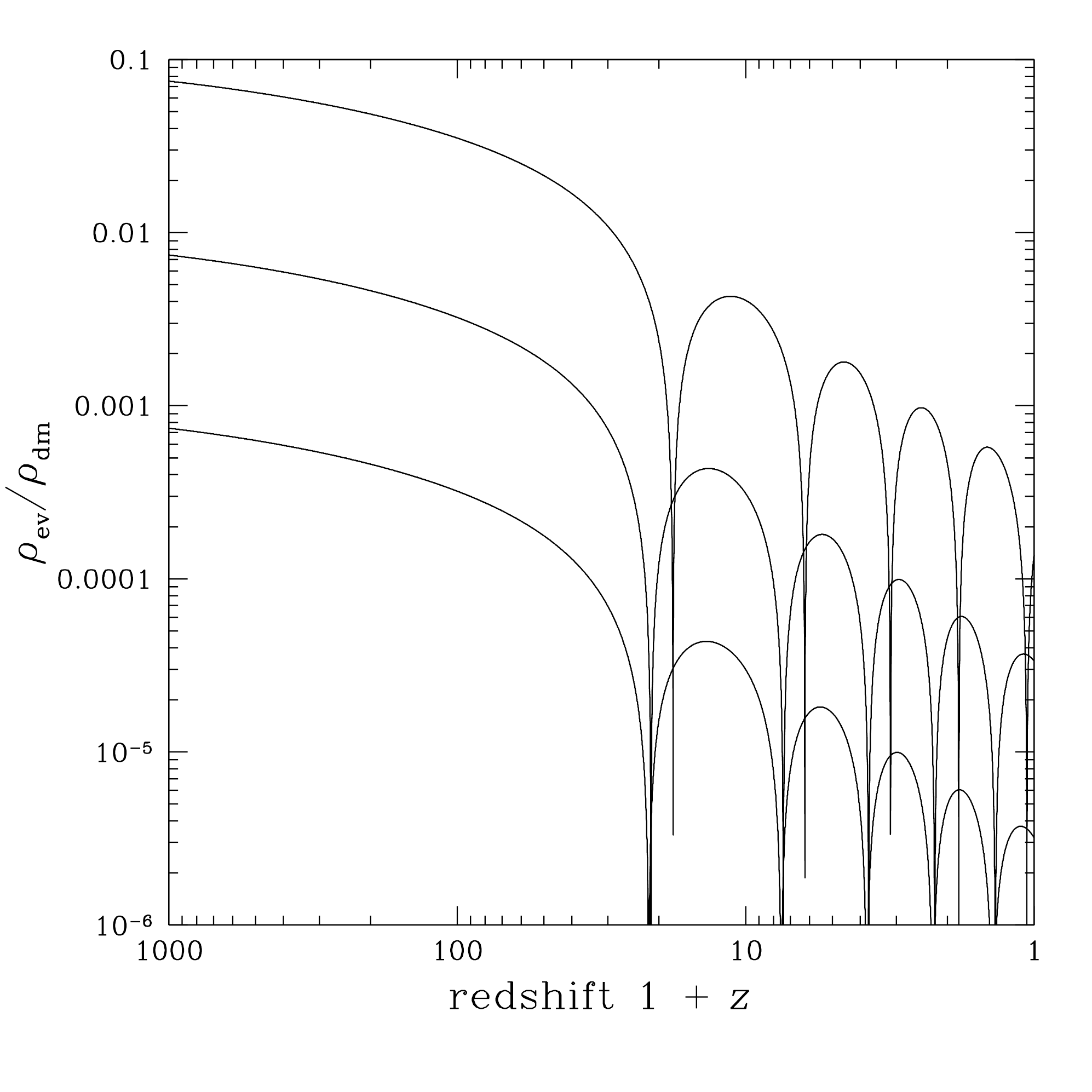} 
\caption{\small Illustration of the evolution of the ratio of evanescent to dark matter mass densities for initial values of the mass  ratio $R_{\rm ev}= 0.1$, 0.01, and 0.001 and the measure of the fifth force in Eq.~(\ref{eq:RF}).\label{Fig:phi}}
\end{center}
\end{figure}

Figure~\ref{Fig:phi} shows the evolution of the evanescent matter mass density from numerical integration of Eqs.~(\ref{eq:pi-iint}) and~(\ref{eq:FLeq}). The product of evanescent matter parameters (Eq.~(\ref{eq:parameters})) is taken to be 
\beq
R_{\rm ev}F_{\rm ev} = 0.216. \label{eq:RF}
\eeq 
This value is chosen so that $\phi$ first passes through zero at redshift $z\simeq 20$, depending slightly on the choice of $R_{\rm ev}$, allowing the possibility of interesting effects on early stages of galaxy formation. The next two subsections illustrate these effects in the approximations of nonlinear spherical symmetry and aspherical linear perturbation theory. Both approximations are quite limited, but but offer preliminary guidance to how the model might affect structure formation within what is allowed by the cosmological tests discussed in Subsection~\ref{Cosmological-tests}. 

\subsection{Spherical accretion model}\label{sec:accretion}

In spherical symmetry and with the time unit $t_e$ (Eq.~\ref{eq:timeunit}) the equations of motion (Eq.~\ref{eq:particle-eqom}) are 
\beqa
&&a{d \over dt}a^2{d x_i\over dt} = x_i - {1\over x_i^2}\sum_{x_j\leq x_i}m_j^{\rm dm} 
    + R_{\rm ev}{|\phi|\over\phi_i}\left(x_i - {1\over x_i^2}\sum_{x_l\leq x_i}m_l^{\rm ev}  \right), 
  \nonumber \\
&&\hspace{-8mm}{a\over |\phi|}{d \over dt}a^2|\phi|{d x_k\over dt} =x_k - 
{1\over x_k^2}\sum_{x_j\leq x_k}m_j^{\rm dm} 
    + R_{\rm ev}\left({|\phi|\over\phi_i}+F_{\rm ev}{\phi_i\over|\phi|}\right)\left(x_k - {1\over x_k^2}\sum_{x_l\leq x_k}m_l^{\rm ev}  \right).
\label{eq:spherical-model}
\eeqa
The first term on the right hand side of each equation replaces the condition that the sums in Eq.~(\ref{eq:particle-eqom}) are ordered to vanish when the mass distribution is homogeneous. The constant shell masses $m_i$ are chosen to produce a small initial departure from a homogeneous mass distribution, here the Plummer form 
\beq
\sum_{x_j\leq x_i} m_j = x_i^3\left[1+{\delta_c\over (1+x_i^2)^{3/2}}\right]. \label{eq:initialconds}
\eeq
The mass unit, $4\pi\bar\rho_{\rm dm}a^3/3$, is the dark matter mass within the Plummer radius $x=1$ in a homogeneous mass distribution.

Numerical integration of the equation of motion of an evanescent particle as $\phi$ passes through zero requires special consideration. Suppose the crossing time $\Delta t$ from $\phi_a$ to $-\phi_a$, where the velocity $v_a$ at $\phi_a$ is nonrelativistic, is short enough that that the evolution of the field may be approximated as $\phi/\phi_a=-2t/\Delta t$, and short enough that the particle momentum $p$ is nearly unchanged. Then the particle Lagrangian (Eq.~\ref{eq:lagrangian}) during this short time interval may be approximated as ${\cal L}_{\rm ev}=-|\phi|\sqrt{1-a^2\dot x^2}$, with momentum $p = |\phi|a^2\dot x(1-a^2\dot x^2)^{-1/2}$. At constant $p$ the particle displacement in the time $\Delta t$ is
\beq
a\,\Delta x = v_a\Delta t\,\sinh^{-1} 1/v_a. \label{throughzero}
\eeq
Though it would be simple to use this relation, the even simpler procedure used here computes all displacements as  $v_a\Delta t$ (in trapezoidal approximation). Since $\Delta t$ is small the logarithm produces only a modest error in the time step across $\phi=0$.  

In the present examples the initial density contrast in Eq.~(\ref{eq:initialconds}) is $\delta_c=0.004$ at initial expansion parameter $a=0.01$ (redshift $z=100z_{\rm eq}\sim 3\times 10^5$), and 4000 mass shells are  spaced at initial radii $x_i\propto i^2$ (a spacing trials suggest produces useful resolution) out to $x=4$. The conventional cosmological parameters used hare are Hubble constant $H_o=70$ km~s$^{-1}$~Mpc$^{-1}$, density parameter $\Omega_m=0.27$ in dark matter (with baryons treated as dark matter), zero space curvature, and redshift $a_o=z_{\rm eq}=3230$ at equal mass densities in radiation and dark matter.  The time unit (Eq.~\ref{eq:timeunit}) is $t_e=6.5\times 10^{12}$~s. The physical length unit in Eqs.~(\ref{eq:spherical-model}) and~(\ref{eq:initialconds}) is the Plummer radius at $z_{\rm eq}$ defined by the initial departure from homogeneity. If the physical length unit is changed to the Plummer radius $\ell_{100}$ scaled up to the present epoch, and expressed in units of 100~kpc, then the physical peculiar velocity of a particle with coordinate velocity $\dot x=dx/dt$ at epoch $t$ is 
\beq
v=146  a(t)\dot x\,\ell_{100}\hbox{ km~s}^{-1}.\label{eq:velocities} 
\eeq

\begin{figure}[htpb]
\begin{center}
\includegraphics[angle=0,width=5.in]{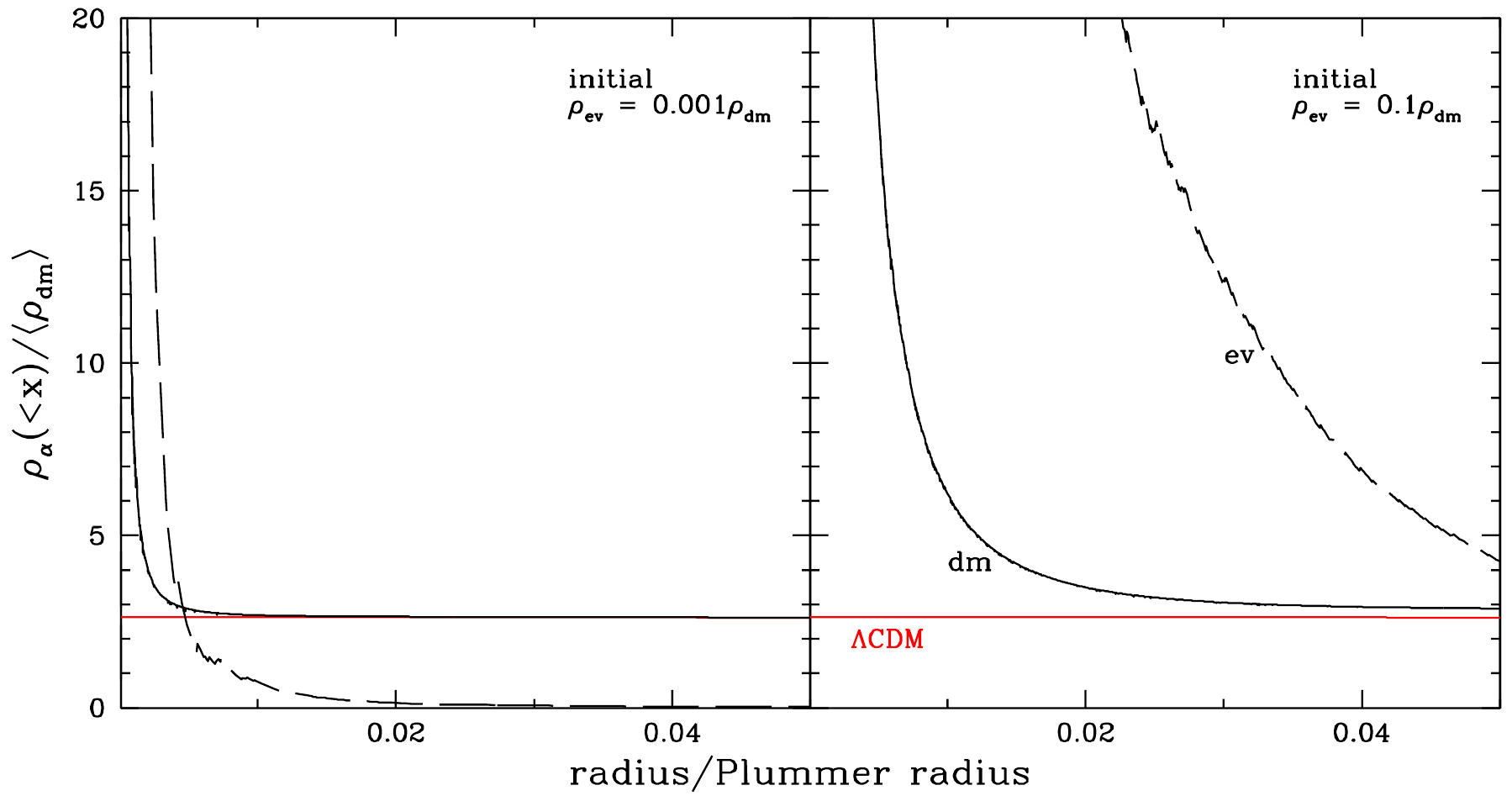}
\caption{\small Mass density runs at redshift $1+z=25$ for dark and evanescent matter plotted as black solid  and dashed curves. The red curve in  the mass density run in $\Lambda$CDM without evanescent matter.  \label{Fig:2} }
\end{center}
\end{figure}

The mass density runs are presented as the value of $\rho_\alpha(<x)$, the mass within $x$, dark or evanescent, divided by the value of the dark matter mass contained within $x$ at the cosmic mean density. Figure \ref{Fig:2} shows the situation at redshift $1+z=25$, somewhat before $\phi$ first passes through zero. The red curve is the mass run without evanescent matter and all other parameters unchanged. Here the central density has grown to 2.6 times the cosmic mean. The curvature of this red line is difficult to see because only the central part of the perturbed mass distribution is shown. The fifth force and the decreasing value of $\phi$ in the evanescent matter both increase the evanescent particle velocities over gravitational free fall, producing the central evanescent mass concentration that gravitationally draws in the smaller spike of dark matter. The mass densities in the spikes near $x=0$ vary as $\rho\sim x^{-2}$ as matter flows radially to and away from $x=0$. At primeval mass ratio $R_{\rm ev}=0.001$ (the left hand panel of Fig.~\ref{Fig:2}) the evanescent mass within radius $0.005$ times the initial Plummer radius has grown to $\sim 3\times 10^{-7}$ times the mass initially within the  Plummer radius, attracting a comparable mass of dark matter.  At $R_{\rm ev}=0.1$ the fifth force is smaller (recall the product $R_{\rm ev}F_{\rm ev}$ in Eq.~\ref{eq:RF} is fixed), but that is offset by the larger gravitational role of the evanescent matter in perturbing the dark matter. Here the masses in evanescent and dark matter within $0.06$ times the Plummer radius are comparable at $\sim 5\times 10^{-4}$ times the mass within the Plummer radius.  Substantial shell crossings at $1+z=25$ and $R_{\rm ev}=0.001$ extend to $x\sim 0.003$ in dark matter and $x\sim1$ in evanescent matter. At  $R_{\rm ev}=0.1$, substantial crossings extend to $x\sim 0.02$ in  dark matter and $x\sim 0.5$ in evanescent matter. 

\begin{figure}[htpb]
\begin{center}
\includegraphics[angle=0,width=5in]{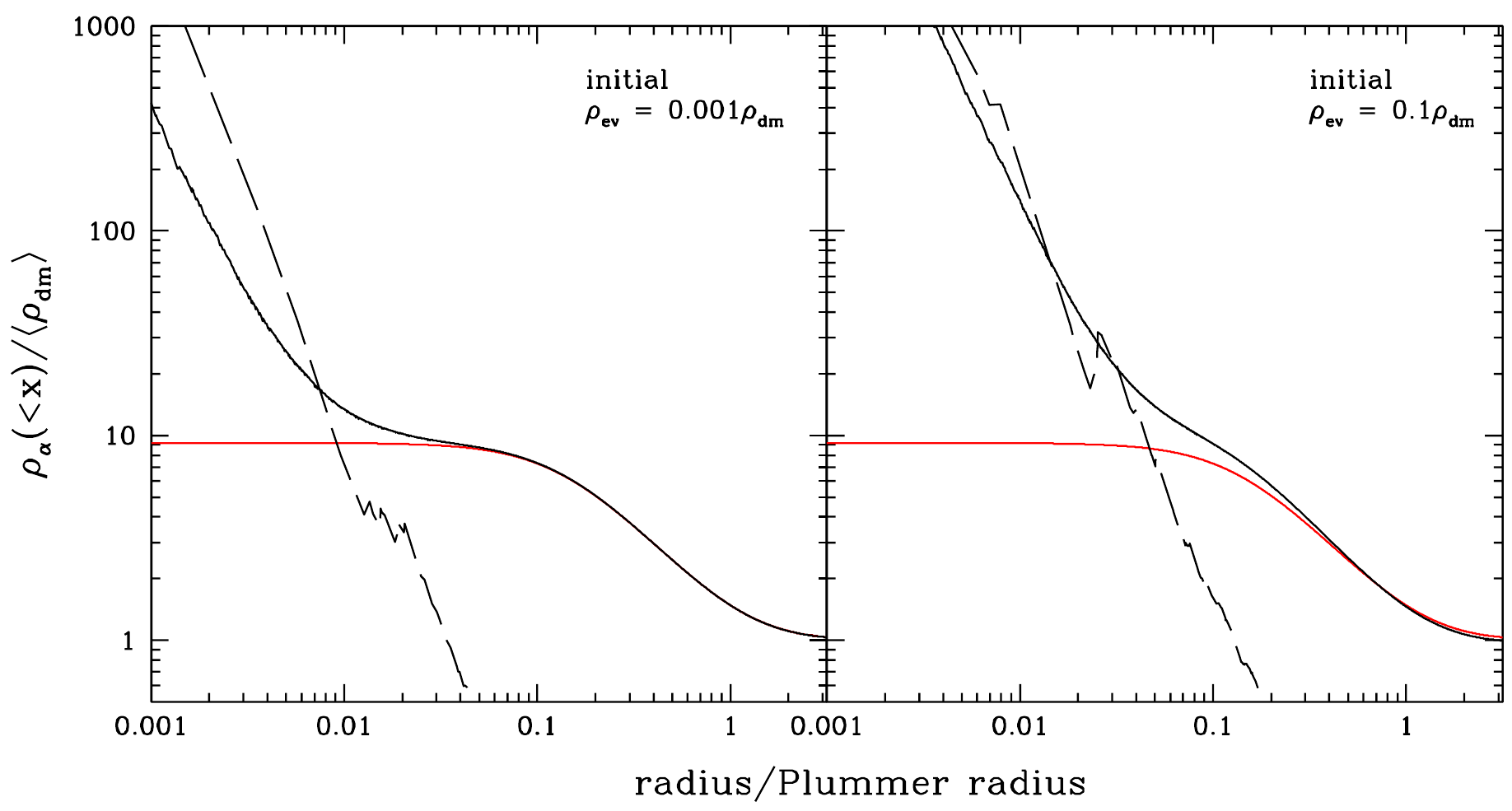}
\caption{\small Mass distributions at $1+z=15$ that evolved from the conditions at $1+z=25$ in  Figure~\ref{Fig:2}.  \label{Fig:3}}
\end{center}
\end{figure}

The mass density runs at $1+z=15$ are shown in Figure~\ref{Fig:3}. When  $\phi$ earlier passed through zero the large evanescent matter shell velocities produced large displacements that left some shells at $1+z=15$ well away from the mass concentration and moving away with speeds $\sim 300\ell_{100}$~km~s$^{-1}$ (Eq.~(\ref {eq:velocities})). At this redshift the shell crossings extend through the full range of evanescent matter shells in the simulation. This means the evanescent mass density is underestimated by the omission of initially more distant shells that would have been at smaller radii at $1+z=15$, but trials with shells at larger initial radii suggest this is not a substantial error.

\begin{figure}[htpb]
\begin{center}
\includegraphics[angle=0,width=5in]{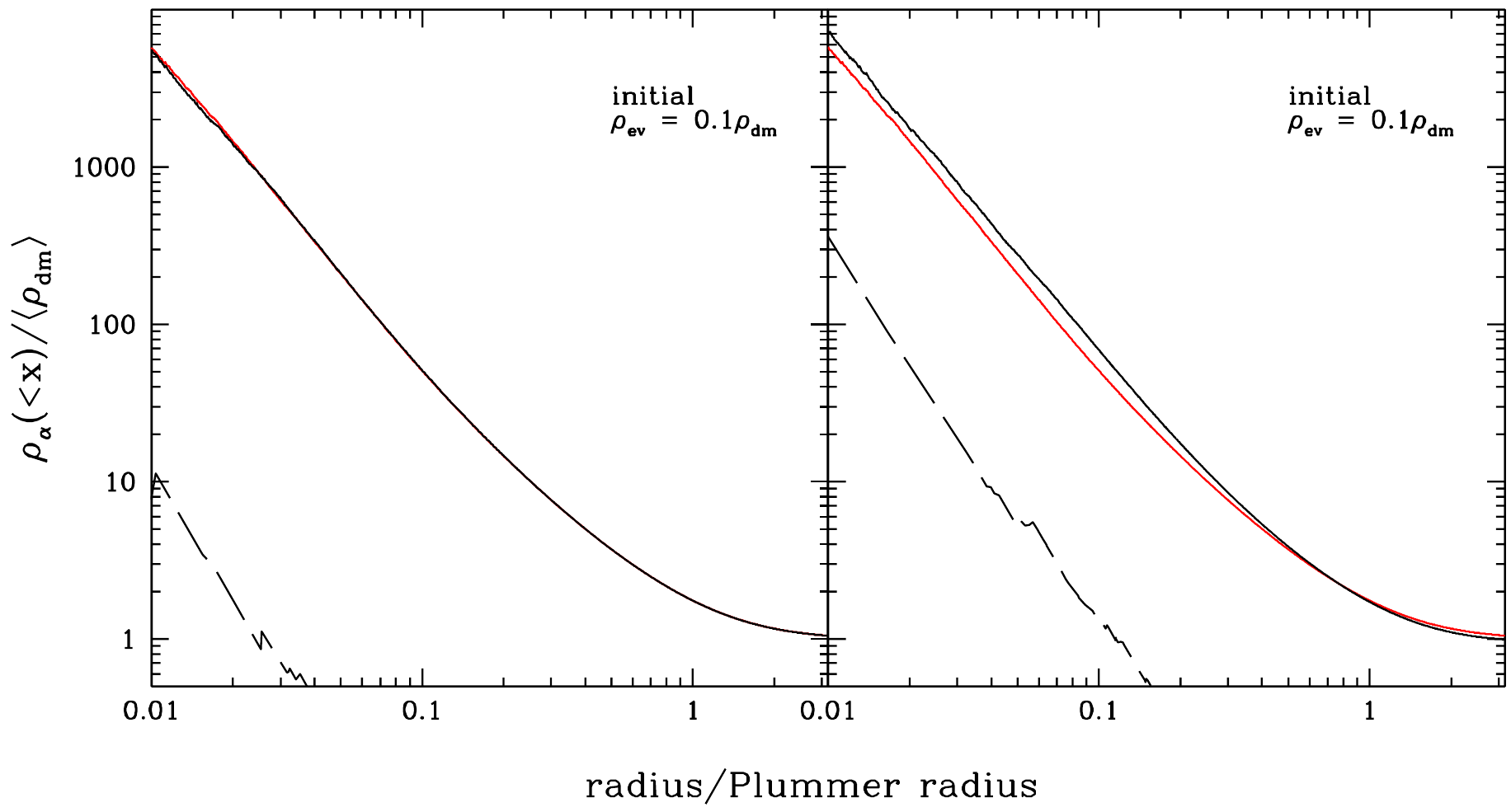}
\caption{\small Mass distributions at $1+z=10$.  \label{Fig:4}}
\end{center}
\end{figure}

At $1+z=10$ (Fig.~\ref{Fig:4}) the evanescent mass is subdominant everywhere, and much of it is streaming away. The dark matter distributions with and without evanescence have grown quite similar despite their different histories. 

\begin{figure}[htpb]
\begin{center}
\includegraphics[angle=0,width=5in]{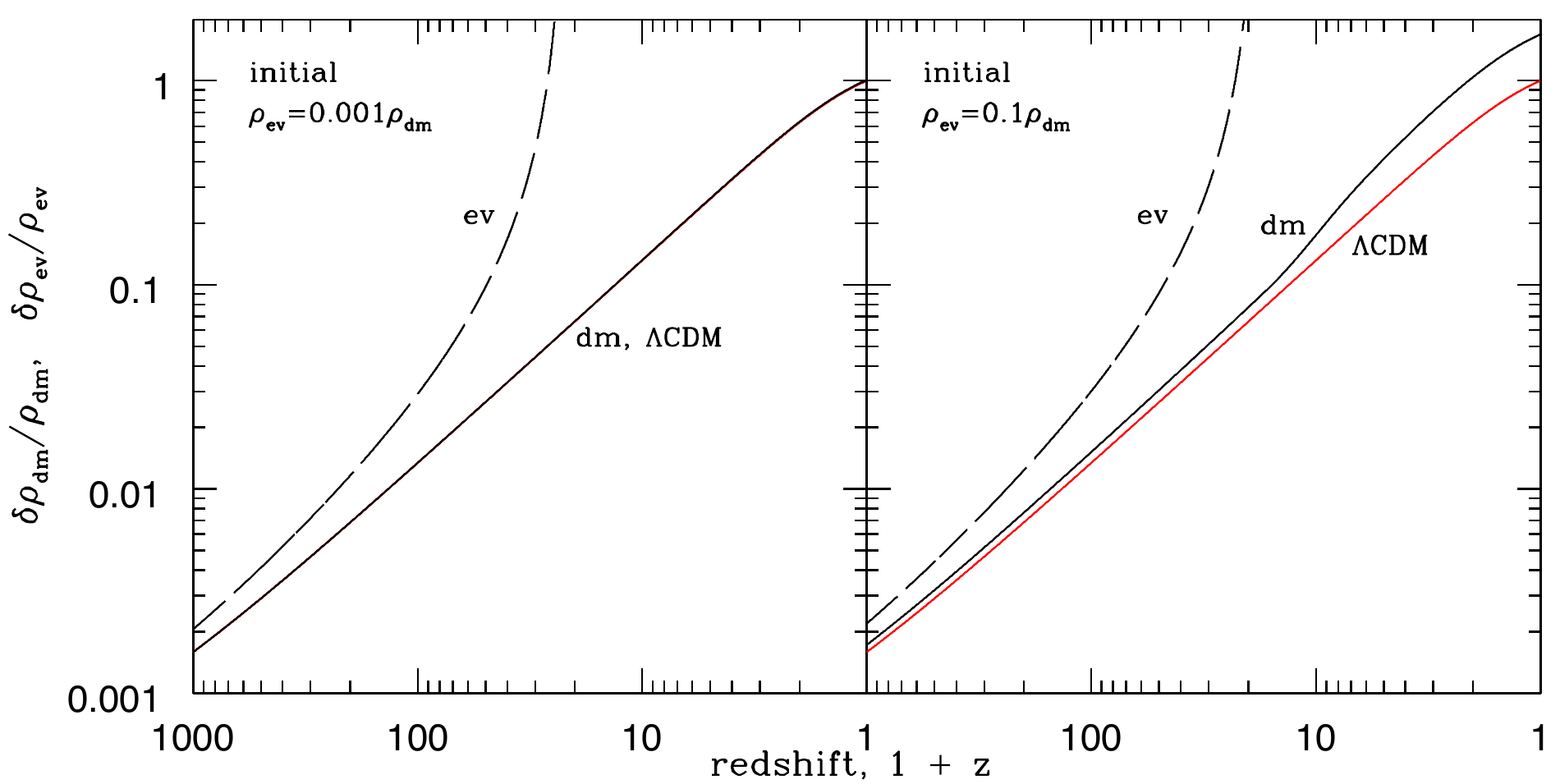}
\caption{\small Evolution of dark and evanescent mass density contrasts in linear perturbation theory.  \label{Fig:5}}
\end{center}
\end{figure}

\subsection{Linear perturbation theory}\label{sec:perturbation}

In linear perturbation theory the density contrast $\delta_\alpha=\delta\rho_\alpha/\rho_\alpha$ and proper peculiar velocity $\vec v_\alpha=a(t)d\vec x_\alpha/dt$ of matter component $\alpha$ satisfy 
\beq
{\partial\delta_\alpha\over\partial t} = -{\nabla\cdot\vec v_\alpha\over a}.
\eeq
The results of multiplying this expression by $a$ or $a|\phi|$, differentiating with respect to time, and using the equation of motion~(Eq.~\ref{eq:eqom}) and the Poisson Eqs.~(\ref{eq:newphifield}) and~(\ref{eq:gravpoisson}) with the conventions in Eqs.~(\ref{eq:parameters}) and~(\ref{eq:timeunit}) are
\beqa
&& {\partial\over\partial t}a^2{\partial\delta_{\rm dm}\over\partial t} = 
{3\over a}\left[\delta_{\rm dm} + R_{\rm ev}{|\phi|\over\phi_i}\delta_{\rm ev}  \right], \nonumber\\
&& \hspace{-8mm}{1\over |\phi |}{\partial\over\partial t} a^2|\phi |{\partial\delta_{\rm ev}\over\partial t} = 
{3\over a}\left[\delta_{\rm dm} + R_{\rm ev}\left({|\phi|\over\phi_i} + 
F_{\rm ev}{\phi_i\over|\phi|} \right) \delta_{\rm ev}  \right] .
\label{eq:perteqns}
\eeqa

The numerical solutions in Figure~\ref{Fig:5} use the same parameters as the spherical model, and treat the evolution of $\delta_{\rm ev}$ through $\phi=0$ as in Eq.~(\ref{throughzero}). As $\phi$ passes through zero $\delta_{\rm ev}$ grows exceedingly large. This is allowed in linear theory but bounded in practice by the nonlinear development of concentrations supported by internal motions. This is taken into account in the numerical solutions by bounding the contrast at $\delta_{\rm ev}=100$. With this crude prescription the present dark matter density contrasts with and without the evanescent component differ by less than the width of the curve for $R_{\rm ev}=0.001$, and at $R_{\rm ev}=0.1$ the evanescent matter makes the dark matter contrast 1.7 times the value without evanescence. 

\subsection{Cosmological tests} \label{Cosmological-tests}

An acceptable adjustment of the standard cosmology must of course pass the established suite of cosmological tests. The analyses in the last section are too schematic to complete checks of consistency with the tests, but they guide a few considerations. 

At the product of parameters in Eq.~(\ref{eq:RF}), and $R_{\rm ev}\leq 0.1$, the evanescent component reduces the expansion time and angular size distance computed from high redshift by less than 1\%. This is  well within the constraints from the cosmological tests. 

The $\Lambda$CDM cosmology predicts that large-scale mass fluctuations normalized to the 3K cosmic microwave background (CMB) anisotropy spectrum have grown to about 80\% of the fluctuations in the present large-scale distribution of $L\la L_\ast$ galaxies. This modest bias seems not unreasonable. Evanescence increases the growth of $\delta_{\rm dm}$, which means it would predict larger present mass fluctuations. If $\delta_{\rm dm}$ were increased by 20\% it would remove the bias, also a not unreasonable situation. A much larger increase would require substantial antibiasing, which is arguably unreasonable. The indication in Figure~\ref{Fig:5} is that $R_{\rm ev}=0.1$ brings the present large-scale mass fluctuation amplitude to about 1.4 times the $\Lambda$CDM prediction, meaning $R_{\rm ev}=0.1$ with Eq.~(\ref{eq:RF}) may exceed the acceptable bound on the model. This is based on a rough approximation to the late time behavior of $\delta_{\rm ev}$, however. A numerical N-body simulation that properly takes account of Eq.~(\ref{throughzero}), as well as the departure from spherical symmetry, is feasible and required for firmer bounds on $R_{\rm ev}$ and $F_{\rm ev}$ from this consideration. 

At high redshift, when plasma and radiation oscillate as a coupled viscous fluid, evanescent matter would act as an addition to the dark matter mass and the fifth force would further increase the growth of mass density fluctuations on scales smaller than the expansion time $t$. The effect would be largest during evolution from $z_{\rm eq}$ to decoupling, as the mass in matter becomes self-gravitating and the largest peak of the CMB anisotropy spectrum forms. Determining whether this can be made to fit the CMB anisotropy spectrum measurements by changing the value of $\Omega_{\rm dm}$ without an unacceptable change in the angular size distance requires a computation in linear perturbation theory for the evolving departures from homogeneity in the distributions of matter and radiation.

The small separations of the solid black and red curves in the spherical model solution at $1+z=10$ in Figure~\ref{Fig:4} agrees with the small separations of $\delta_{\rm dm}$ and $\delta_{\rm ev}$ at $1+z=10$ in the linear perturbation solution in Figure~\ref{Fig:5}. That is, within the range of parameters considered here one may expect little effect on the present masses of galaxies and clusters of galaxies, though one would look for earlier assembly and relaxation of central nonlinear concentrations, as in Figure~\ref{Fig:3}, depending on when $\phi$ first passes through a minimum.

\section{Discussion}\label{sec:remarks}

Section~\ref{sec:phenomenology} reviews several challenges to the standard cosmology that at least some experts consider serious. It may prove to be possible to reconcile all these challenges to $\Lambda$CDM, but that path seems tortuous enough to motivate consideration of how the cosmology might be adjusted to present an apparently more natural approximation to what is observed. I have stressed that a common feature of the  challenges is that galaxies might be more readily understandable if some process caused them to be assembled earlier than expected in the current paradigm, with resulting suppression of merging and accretion at lower redshifts. The illustrations in Section~\ref{AnIllustration} show how evanescent matter can make this happen. 

The evanescent matter is defined by the Lagrangian density in Eq.~(\ref{eq:action}), which is to be added to  $\Lambda$CDM. As noted in Section~\ref{sec:ev}, superstring theory offers the scalar or pseudoscalar fields that enter this proposed Lagrangian, and the standard model for particle physics offers precedence for the coupling of these fields to the spin-1/2 evanescent matter particle. This particle may be chiral, as in the standard model, but the highly schematic variant  in Eq.~(\ref{eq:action}) lacks gauge fields and a Higgs mechanism. The very special parameter choice in Eq.~(\ref{eq:RF}) is required to make the evanescent matter do something significant and not manifestly unacceptable. There is ample precedent for special parameter values, of course, in $\Lambda$CDM most notably the curiously value of Einstein's $\Lambda$. 

My conclusion from the considerations in the preceding two paragraphs is that the evanescent matter model offers an interesting direction to consider for the purpose of exploring possible implications of some very interesting challenges to $\Lambda$CDM. The illustrative examples in Section~\ref{AnIllustration} suggest that in this picture the evanescent matter mass density may be subdominant except where the fifth force and decreasing particle mass drive early formation of local strongly nonlinear concentrations of evanescent matter that gravitationally attract dark matter concentrations earlier than expected in standard $\Lambda$CDM. With the parameters in this illustration the evanescent matter concentrations are transient, dispersed by the increasing particle velocities as the particle  mass decreases. Not yet explored even at the simple level of Section~\ref{AnIllustration} is the version in Eq.~(\ref{eq:classical-2-particle-acton}) with two fields and $\lambda_1\not=\lambda_2$. This can eliminate the zeros of the evanescent matter mass (because $\phi_1$ and $\phi_2$ are not likely to pass through zero at the same time). The resulting elimination of relativistic motion would allow evanescent mass concentrations to last longer, increasing their gravitational effect on the dark matter, which may offer a better model for early galaxy formation. Beyond the two-field model one can invent still more baroque generalizations of the evanescent matter picture, but that discussion might await further exploration of the one- and two-field models. 

Another model motivated by the same phenomenology adds a fifth force to the $\Lambda$CDM dark matter rather than to an added evanescent component \cite{FarrarPeebles,PeebGub}. The idea of a fifth force on the dark matter is challenged by the observation of the leading and trailing streams of the Sagittarius satellite, which are characteristic of normal gravitational destruction rather than the effect of a fifth force \cite{KesdenKamionkowski,Kesden}, though the dark matter fifth force may be viable if the Sagittarius galaxy were initially massive and is now suffering destruction of the central baryon-dominated part \cite{Keselman}. But in either case this constraint is not relevant for evanescent matter now scattered outside dark matter halos.

A full-scale numerical N-body simulation likely is needed to explore what are now speculative ideas about how the   evanescent matter model could affect the growth of cosmic structure. Since a growing dark matter halo consists of merging subhalos, one might expect to see that some subhalos are dense enough as the evanescent matter mass approaches zero to have grown into transient tight concentrations that gravitationally attract tight dark matter concentrations, replacing the single spike of dark matter in the spherical model in Section~\ref{AnIllustration} with spikes scattered through the denser parts of the halo.  One might also expect that these mass concentrations of dark matter merge as the halo relaxes, producing a single dark matter concentration that might be similar to the illustration in Figure~\ref{Fig:3}, though almost certainly with smaller mass for given evanescence parameters. In more massive halos this central mass concentration might promote formation of massive black holes in young galaxies. I note in passing that that could help account for the presence of luminous quasars apparently powered by black holes with masses $\ga 10^9$ Solar masses at redshift $z\simeq 7$ \cite{Mortlock}. The merging of spikes might be expected to have affected star formation in a young galaxy. It is easy to speculate, but difficult to check, that the merging enhances star formation, perhaps enough to promote early formation of stellar bulges before large-scale structure had grown enough to produce significant variations in the environments of protogalaxies. This is in the direction suggested by point (2) in Section~\ref{sec:phenomenology}, the strikingly modest sensitivity of elliptical galaxies with given stellar velocity dispersions to their present environments. One  might also imagine that the formation of tight spikes of evanescent matter requires that the mass density in a protogalaxy exceeds a critical value. A threshold might help explain why some large spiral galaxies end up with classical bulges in while in a pure disk galaxy star formation awaited the settling of diffuse baryons onto the growing disk (point (1) in Sec.~\ref{sec:phenomenology}). Even more speculative, but worth considering, is the idea that the dispersal of the evanescent matter concentrations may help account for the apparent expansion of large early-type galaxies (point~(3)). Speculation on point~(4) is best confined to the remark that evanescent matter would affect early formation of structure on small scales in ways to be explored in numerical simulations. 

Finally, I believe it is well to bear in mind that natural science generally makes progress by successive approximations. One might accordingly suspect that $\Lambda$CDM is a good approximation rather than a final theory, and that a more accurate theory of the dark sector might be expected to share to some degree the complexity of physics in the visible sector. (This is a thought I find expressed more frequently in conversations than in print.) If there is a better cosmology we may be led to it by anomalies within $\Lambda$CDM. The evanescent matter model adds to the illustrations how consideration of the evidence, which has grown quite serious, can motivate viable and observationally interesting adjustments of the current paradigm in cosmology. 
 
\begin{acknowledgement}
I have benefitted from discussions with Michael Blanton, John Kormendy, Yen-Ting Lin, Piero Madau, Chris McKee, Ben Moore, Surhud More, John Moustakas, and Gunagtun Ben Zhu, and from comments from the referees. 

\end{acknowledgement}

\def\bstname{adp}

\end{document}